\documentclass[aps,prx,longbibliography,twocolumn,showpacs,noeprint]{revtex4-2}
\usepackage{array}[=2016-10-06]

\usepackage{CJK}

\usepackage{graphicx}

\usepackage{amsmath}
\usepackage{dcolumn}

\usepackage{bm}
\usepackage[normalem]{ulem}
\usepackage{color}
\usepackage{hyperref}

\usepackage{epstopdf}
\newcommand{\bew}{\begin{widetext}}
\newcommand{\ew}{\end{widetext}}

\newcommand{\ii}{{\rm i}}
\newcommand{\bx}{\mathbf{x}}

\newcommand{\bq}{\mathbf{q}}

\newcommand{\br}{\mathbf{r}}

\newcommand{\bff}{\mathbf{f}}

\newcommand{\bg}{\mathbf{g}}

\newcommand{\sep}{ \ \ \ , \ \ \ }

\newcommand{\beq}{\begin{equation}}
\newcommand{\eeq}{\end{equation}}
\newcommand{\beqn}{\begin{eqnarray}}
\newcommand{\eeqn}{\end{eqnarray}}

\newcommand{\tand}{\text{and}}

\allowdisplaybreaks[1]

\begin{document}

\title{Diversity of critical phenomena in the ordered phase of polar active fluids}
\author{Patrick Jentsch}
	\email{patrick.jentsch@embl.de}
	\affiliation{Cell Biology and Biophysics Unit, European Molecular Biology Laboratory Heidelberg, Meyerhofstrasse 1, 69117 Heidelberg, Germany}
	\author{Chiu Fan Lee}
	\email{c.lee@imperial.ac.uk}
	\affiliation{Department of Bioengineering, Imperial College London, South Kensington Campus, London SW7 2AZ, U.K.}
	\begin{abstract}
We present a comprehensive analytical linear stability analysis of the Toner–Tu model for polar active fluids in the ordered phase.
Our results provide exact instability criteria and demonstrate that all generic hydrodynamic instabilities fall into two fundamental categories, distinguished by their scaling with the wavevector magnitude.
By applying a general criticality condition, we show that each instability can give rise to a critical point by fine-tuning only two parameters.
We identify four previously unreported critical points of the Toner–Tu model, two of which already display nonequilibrium critical behavior that extends beyond known universality classes at the linear level. We further construct explicit hydrodynamic models that realize each newly identified critical point, establishing their physical attainability and providing concrete targets for future renormalization-group analyses and microscopic model studies. Altogether, our framework offers a unified theoretical foundation and a practical roadmap for the systematic discovery of new universality classes in active matter.
\end{abstract}

\maketitle

\section{Introduction}
	Active matter physics governs the behavior of nonequilibrium systems whose constituents can exert forces on each other and on their surroundings \cite{marchetti_rmp13}. The rise of active matter physics stems not only from its direct relevance to biology but also from its role as a fertile ground for discovering novel physical phenomena—most notably, the many new universality classes (UCs) uncovered over the past decade \cite{chen_njp15,gelimson_prl15,toner_prl18,chen_prl20,mahdisoltani_prr21,zinati_pre22,chen_prl22a,jentsch_prr23,
	cavagna_natphys23,vanderkolk_prl23,besse_prl23,toner_pre24,chen_pre24,miller_pre24,jentsch_prl24,wong_a25, legrand_a25}.
Most of these UCs have been identified through renormalization group (RG) analyses of the celebrated Toner–Tu (TT) model \cite{toner_prl95,toner_pre98}, typically explored in various limiting regimes. The TT model describes a generic dry polar active fluid and, due to its inherent complexity, continues to conceal much of its rich physics.

Even focusing solely on critical behavior, it was initially unclear whether the TT model could exhibit an order–disorder critical transition at all. This uncertainty partly arose from the microscopic Vicsek model of flocking \cite{vicsek_prl95}, which directly inspired the TT framework \cite{toner_b24}. The Vicsek model shows no continuous order–disorder transition, as such a transition is always preempted by a discontinuous one leading to the banding regime—a phase-separated coexistence of ordered (moving) and disordered (non-moving) states \cite{gregoire_prl04,bertin_pre06,solon_prl15}.
However, the TT model, as a generic hydrodynamic theory of polar active fluids, is not constrained by the phenomenology of the microscopic Vicsek model. 
Indeed, the existence of a critical order–disorder transition was recently demonstrated \cite{nesbitt_njp21} for systems of polar active fluids whose collective motion ceases at high density, such as motile agents exhibiting contact inhibition of locomotion in certain cellular tissues \cite{schnyder_scirep17}. This transition can be accessed by fine-tuning two model parameters, rendering it no less experimentally relevant than, for instance, thermal gas–liquid phase separation.
Yet, due to the TT model’s complexity, the RG fixed point governing the scaling behavior of this critical point in physical dimensions remains analytically unresolved. A multicritical extension of the model—obtained by tuning four parameters—proved more tractable, revealing, perhaps unsurprisingly, a distinct nonequilibrium universality class \cite{jentsch_prr23}.

Intriguingly, Miller and Toner \cite{miller_prl24} recently showed that phase-transitions in the TT model are not limited to the order–disorder transition but can also arise within the ordered phase itself. They demonstrated that the “compressibility” of an active, chemotactic system can become negative, leading to a phase separation between two moving phases with distinct speeds \cite{miller_prl24,miller_pre24b}. Unlike previously reported coexistence between moving phases separated by an interface perpendicular to the direction of motion \cite{bertrand_prr22}, here the interface lies parallel to it. Moreover, Miller and Toner identified a special point on the boundary of this instability region—analogous to the critical point in thermal gas–liquid separation—whose scaling behavior defines yet another new universality class via perturbative RG analysis \cite{miller_pre24}.

In this work, we perform a comprehensive linear stability analysis of the TT model’s ordered phase and derive analytical criteria for instability. We uncover two generic classes of instabilities, distinguished by their scaling with wavevector magnitude. Furthermore, we identify four critical points, so far unreported for the TT model, and characterize their scaling behavior at the linear level.
Two of these exhibit novel nonequilibrium critical behaviors beyond known universality classes.
Finally, we construct explicit hydrodynamic models, grounded on distinct physical mechanisms, capable of exhibiting these novel types of criticality.

\section{Linear stability of the TT ordered phase}
\label{sec:instability}
Derived solely from symmetry considerations and conservation laws, 
the Toner–Tu (TT) model provides a generic hydrodynamic description of dry polar active fluids, 
analogous to how the Navier–Stokes equations describe simple fluids \cite{toner_b24}. 
In terms of the hydrodynamic variables---the mass density field $\rho$ 
and the momentum density field $\bg$---the equations of motion read
\begin{equation}
	\label{eq:cont}
	\partial_t \rho + \nabla \!\cdot\! \bg = 0 \, .
\end{equation}
\begin{widetext}
\begin{align}
	\label{eq:tt}
	\nonumber
	\partial_t g_i 
	&+ \lambda_1 g_j \nabla_j g_i 
	+ \lambda_2 g_i \nabla_j g_j
	+ \lambda_3 g_j \nabla_i g_j
	= -U g_i 
	- \kappa_1 \nabla_i \rho  
	- \nu_1 g_i g_j \nabla_j \rho \\[3pt]
	&\quad 
	+ \mu_1 \nabla^2 g_i 
	+ \mu_2 \nabla_i \nabla_j g_j
	+ \mu_3 g_j g_k \nabla_j \nabla_k g_i 
	+ \mu_4 g_i g_j \nabla^2 g_j
	+ \mu_5 g_i g_j \nabla_j \nabla_k g_k
	+ \mu_6 g_j g_k \nabla_k \nabla_i g_j 
	+ \mathrm{h.o.t.} + \bff \, ,
\end{align}
\end{widetext}
where summation over indices is implied and all coefficients $\lambda_{a}$, $U$, $\kappa_1$, $\nu_1$, and $\mu_{a}$
generally depend on the scalar quantities $|\bg|^2$ and $\rho$; 
``h.o.t.'' denotes higher–order derivative terms.

We next examine the {\it linear stability} of the ordered phase by expanding 
the hydrodynamic fields around the homogeneous, collectively moving state:
\begin{equation}
\rho(t,\br) = \rho_0 + \delta\rho(t,\br) \sep 
\bg(t,\br) = \bg_0 + \bg_x(t,\br) + \bg_\bot(t,\br) \, ,
\end{equation}
where, without loss of generality, $\hat{\bx}$ denotes the spontaneously chosen flocking direction, 
and $\bg_0 = \langle \bg \rangle = g_0 \hat{\bx}$ is the mean momentum density. 
The components $\bg_x = \hat{\bx} (\bg \!\cdot\! \hat{\bx} - g_0) \equiv g_x \hat{\bx}$ 
and $\bg_\bot = \bg - \bg_0 - \bg_x$ represent the parallel and perpendicular deviations 
of the momentum field with respect to $\hat{\bx}$. 
The linearized TT equations of motion then read \cite{SM}:
\begin{align}
\label{eq:linear_eom_1}
\partial_t \delta\rho &+ \nabla_\bot\!\cdot\!\bg_\bot + \partial_x g_x = 0 \, , \\[3pt]
\label{eq:linear_eom_2}
\nonumber
\partial_t \bg_x 
&+ \lambda_{\rm tot} g_0 \partial_x \bg_x 
+ \lambda_2 \bg_0 \nabla_\bot\!\cdot\!\bg_\bot
= -\beta g_0^2 \bg_x - \alpha_1 \bg_0 \delta\rho \\[2pt]
\nonumber
&\quad 
- \kappa_x \hat{\bx} \partial_x \delta\rho
+ (\mu_\perp^{xx} \nabla_\bot^2 + \mu_x^{xx} \partial_x^2) \bg_x \\[2pt]
&\quad 
+ \mu^{xL} \hat{\bx} \partial_x \nabla_\bot\!\cdot\!\bg_\bot \, , \\[4pt]
\nonumber
\label{eq:linear_eom_3}
\partial_t \bg_\bot 
&+ \lambda_1 g_0 \partial_x \bg_\bot 
+ \lambda_3 g_0 \nabla_\bot g_x
= (\mu_1 \nabla_\bot^2 + \mu_x \partial_x^2) \bg_\bot \\[2pt]
&\quad 
- \kappa_1 \nabla_\bot \delta\rho 
+ \mu_2 \nabla_\bot (\nabla_\bot\!\cdot\!\bg_\bot)
+ \mu^{Lx} \nabla_\bot (\partial_x \bg_x) \, .
\end{align}
All couplings are evaluated at the background values $(\rho_0, \bg_0)$; 
their relations to those in Eq.~\eqref{eq:tt} are detailed in \cite{SM}.

The linear stability analysis amounts to solving for the temporal eigenvalues 
of this system in Laplace–Fourier space 
(Laplace in time, Fourier in space), i.e., 
$\delta\rho(t,\br) = \delta\rho \, e^{s t - \ii \bq \cdot \br}$, 
and analogously for $\bg_x$ and $\bg_\bot$. 
We analyze this eigenvalue problem in the small–wavenumber (hydrodynamic) limit; 
see \cite{SM} for details.

For a $d$–dimensional system there are $(d+1)$ eigenvalues. 
Besides the eigenvalue
\beq
\label{eq:C0}
C_0 \equiv -\beta g_0^2 \, ,
\eeq
to leading order in $q$, which is necessarily negative in the ordered phase, 
there exist $(d-2)$ degenerate eigenvalues associated with the transverse Goldstone modes,
\begin{equation}
  \label{eq:instab_goldstone}
  E_T = -\ii \lambda_1 g_0 q_x - \mu_x q_x^2 - \mu_1 q_\perp^2 \, ,
\end{equation}
and two additional eigenvalues $E_\pm$, which, to order $\mathcal{O}(q^{ 2})$, are given by
\beq
\label{eq:evs}
E_\pm = \ii q A_1(\theta) 
\pm \ii q \sqrt{A_2(\theta)} 
- q^2 B_1(\theta) 
\pm q^2 \frac{B_2(\theta)}{\sqrt{A_2(\theta)}} \, ,
\eeq
where the wavevector components are expressed in polar coordinates 
relative to the $x$–axis: $q_x = q \cos\theta$ and $q_\perp = q \sin\theta$. 
The functions $A_i(\theta)$ and $B_i(\theta)$ are real functions of $\theta$ 
that also depend on the model parameters; 
their analytical expressions are provided in \cite{SM}. 
As expected, these eigenvalues are consistent with the results of Ref.~\cite{toner_pre12}.

\subsection{Two types of instabilities}
\label{sec:C1to4}
We now investigate the stability of the ordered phase by determining the conditions under which the real parts of the eigenvalues become positive. 
From the form of Eqs.~\eqref{eq:instab_goldstone} and \eqref{eq:evs}, linear instabilities can be classified into two generic types according to their scaling with wavenumber $q$:  
(i) {\it Type I} instabilities, which scale as $\mathcal{O}(q)$; and  
(ii) {\it Type II} instabilities, which scale as $\mathcal{O}(q^2)$.

Since $\lambda_1 g_0$ is always real, a {\it Type I} instability can only occur in the $E_\pm$ modes \eqref{eq:evs}, which happens
when $A_2(\theta)$ becomes negative for at least some $\theta$, rendering its square root imaginary.  
Explicitly, we find \cite{SM}
\begin{equation}
	A_2(\theta) 
	= \frac{\kappa_1 \beta - \alpha_1 \lambda_3}{\beta} \sin^2\!\theta 
	+ \left(\frac{\alpha_1 + \lambda_1 \beta g_0^2}{2 \beta g_0}\right)^2 \!\cos^2\!\theta \, .
\end{equation}
Only the first term can be negative, implying that the most unstable wavevector for a {\it Type I} instability is {\it always} oriented perpendicular to the flocking direction ($\theta = \pi/2$).  
Due to the symmetry of $A_2$ , we can restrict our analysis to $\theta \in [0,\pi/2]$.  
The {\it Type I} instability thus occurs when $A_2(\pi/2) < 0$, corresponding to the condition
\beq
\label{eq:type1}
C_1 \equiv \alpha_1 \lambda_3 - \beta \kappa_1 > 0 \, ,
\eeq
which reproduces the instability criterion of Ref.~\cite{miller_pre24b}.  

{\it Type II} instabilities, scaling as $\mathcal{O}(q^2)$, are subdominant to {\it Type I} and therefore appear only when the latter are absent (i.e., when $A_2(\pi/2) > 0$).  
Focusing first on the transverse Goldstone modes with eigenvalues $E_T$ \eqref{eq:instab_goldstone}, 
{\it Type II} instabilities occur when the most unstable wavevectors point either  
perpendicular to the flocking direction, if
\begin{equation}
\label{eq:C3}
	C_2 \equiv  -\mu_1 > 0 \sep \tand \sep \mu_1 < \mu_x \, ,
\end{equation}
or parallel to it, if
\begin{equation}
\label{eq:C2}
	C_3 \equiv -\mu_x > 0 \sep \tand \sep \mu_x < \mu_1 \, .
\end{equation}
In the former case, when $d > 2$, there exist infinitely many equally unstable directions, a feature that will have important implications for the scaling behavior discussed below.

Alternatively, Type II instabilities can also arise in the $E_\pm$ modes \eqref{eq:evs} when
\beq
\label{eq:type2}
C_4 \equiv 
\sup_{\theta \in [0, \pi/2]} 
\left[ 
	- B_1(\theta) 
	+ \left| \frac{B_2(\theta)}{\sqrt{|A_2(\theta)|}} \right| 
\right] > 0 \, .
\eeq
Again, the range of $\theta$ can be restricted to $[0,\pi/2]$ since the argument of the supremum is symmetric under reflection through the $\hat{\bx}$ and $\hat{\bq}_\bot$ axes, i.e., under $\theta \to \pi - \theta$ and $\theta \to -\theta$ \cite{SM}. Condition $C_4$ picks out the most unstable of the two modes $E_\pm$. For $\theta = 0$, $E_-$ coincides with $E_T$. When $E_-(\theta=0)$ is the most unstable mode, $E_T$ is thus equally unstable and both the $C_4$ and $C_3$ instabilities are triggered at the same time \cite{SM}. At first glance, this might seem surprising. But for $\bq=\bq_x$, $\hat \bq_\bot$ is no longer a special direction and $\bg_L$ is no longer well defined. Instead $\bg_T$ now has an additional direction, which is reflected by the Eigenvalues of $\bg_L$ merging with those of $\bg_T$ in this limit.

The classification into these two instability types, together with the analytic expressions 
for all instabilities up to $\mathcal{O}(q^2)$ in Eqs.~\eqref{eq:instab_goldstone}, \eqref{eq:type1}, and \eqref{eq:type2}, 
constitutes the first key result of this paper.  

Finally, applying our analysis to the hydrodynamic equations of the Vicsek model derived via the Boltzmann–Ginzburg–Landau approach 
\cite{bertin_pre06,bertin_jpa09,peshkov_epjst14} reproduces previous numerical stability results \cite{peshkov_epjst14}, 
thus further validating our analytical framework.

\subsection{Phase separation vs.~pattern formation}
The nature of the steady state within the unstable regimes cannot be determined from linear stability analysis alone \cite{cross_b09}. 
In systems with conserved mass, the onset of a linear instability in the $q \!\to\! 0$ limit is often interpreted as an indication of phase separation. 
Indeed, such behavior has been reported in several active matter models \cite{nesbitt_njp21,bertrand_prr22,miller_pre24}. 
However, other models—most notably the Vicsek model—exhibit pattern-forming states with smectic \cite{solon_prl15} 
or cross-sea–like \cite{chate_annrev20,kuersten_prl20} order instead. 
In particular, we speculate that the recently described cross-sea phase \cite{chate_annrev20,kuersten_prl20} 
may be associated with the finite–$\theta$ instabilities identified near the order–disorder transition 
(cf.~Eq.~\eqref{eq:type2}; see also Ref.~\cite{peshkov_epjst14}).

\section{Criticality criteria}
Regardless of the eventual steady state reached in the unstable regime 
(e.g., whether phase separated or patterned), 
the transition from a spatially homogeneous state into that regime 
corresponds generically to a phase transition. 
Based on a mean-field analysis, such a transition would appear to be 
{\it critical} (i.e., continuous) whenever the condition 
$C_i(\rho_0) = 0$ (for one of $i=\{0,\dots,5\}$) is satisfied. 
However, this is {\it generally not the case}. 
When the system lies precisely at the stability–instability boundary, 
local fluctuations of the hydrodynamic fields ($\rho$ and $\bg$ here) 
drive parts of the system into the unstable regime, 
amplifying these fluctuations 
(see, e.g., Refs.~\cite{lee_pre10,martin_prl21} for polar active fluids). 
Consequently, the homogeneous state is intrinsically unstable 
exactly at the boundary.

In thermal phase separation, this manifests as the appearance of 
additional instability regions in the phase diagram—the nucleation and growth regimes—
flanking the spinodal region \cite{weber_rpp19}. 
A similar phenomenology occurs in polar active fluids 
\cite{nesbitt_njp21,bertrand_prr22,miller_prl24}. 
As a result, the transitions associated with these instabilities 
are generically {\it discontinuous}.

Since our focus here is on {\it critical} transitions, 
the natural question is how such continuous transitions can emerge 
from these typically discontinuous ones. 
Dissecting the argument above, a continuous transition can occur 
if the system is tuned to the stability boundary 
{\it and} local fluctuations do not drive it into the unstable regime. 
Because we are considering the ordered phase of polar active fluids, 
the momentum field $\bg$ is generically stable, 
allowing us to focus on the density field alone. 
This leads to two simultaneous conditions for criticality:
\begin{equation}
	\label{eq:crit_cond}
	C_i(\rho_0) = 0 
	\quad \text{and} \quad 
	\frac{\partial C_i(\rho_0)}{\partial \rho_0} = 0 \, ,
\end{equation}
for $i = \{0,\dots,5\}$. 
These are precisely the criteria used to locate critical points 
in both equilibrium systems \cite{hohenberg_rmp77} 
and active phase-separating systems 
\cite{cates_annrev15,nesbitt_njp21,bertrand_prr22,jentsch_prr23,miller_prl24,wong_a25}, 
although this connection has not always been made explicit. 
Beyond phase separation, the same criteria~\eqref{eq:crit_cond} 
also identify the multicritical Lifshitz point 
that separates homogeneous and patterned states.

Because the criticality condition requires two simultaneous constraints, 
achieving criticality generically demands fine-tuning of only two control parameters.  
At such a critical point, the system becomes invariant under rescaling:
\beq
\label{eq:rescale}
\br \!\to\! \br e^{\ell}, \ \
x \!\to\! x e^{\zeta \ell}, \ \
t \!\to\! t e^{z \ell}, \ \
\rho \!\to\! \rho e^{\chi_\rho \ell}, \ \
\bg_\gamma \!\to\! \bg_\gamma e^{\chi_\gamma \ell} \, ,
\eeq
where $\gamma = x, L, T, \ldots$ denotes the relevant components of $\bg$.  
The associated critical exponents 
$\{z, \zeta, \chi_\rho, \chi_\gamma\}$ are universal 
and hence define the universality class of the transition.

The criteria~\eqref{eq:crit_cond} apply to all five instability conditions 
discussed above—Eqs.~\eqref{eq:C0}, \eqref{eq:type1}–\eqref{eq:type2}. 
Since the critical points associated with $C_0$ and $C_1$ 
have already been analyzed in 
Refs.~\cite{nesbitt_njp21,bertrand_prr22,miller_pre24}, 
we focus here on the critical instabilities corresponding to 
$C_2$, $C_3$, and $C_4$, which we identify and characterize for the first time.

\section{Critical points from $C_2$ and $C_3$}
\label{sec:C23}
Applying the criticality criteria~\eqref{eq:crit_cond} to $C_2$ and $C_3$ yields, 
under the assumption that no other instability intervenes, 
critical behavior within the $(d\!-\!2)$ transverse Goldstone modes 
(denoted $\bg_T \!\equiv\! \bg_\bot - \hat{\bq}_\bot (\hat{\bq}_\bot \!\cdot\! \bg_\bot)$) \cite{SM}. 
At the linear level, these transverse modes decouple completely from all others. 
After transforming to the comoving frame, $\br \!\to\! \br + \lambda g_0 t$, 
their linearized equation of motion reads
\begin{equation}
	\partial_t \bg_T 
	= \mu_x \partial_x^2 \bg_T 
	+ \mu_1 \nabla_\bot^2 \bg_T 
	- \nu \nabla^4 \bg_T \, ,
\end{equation}
where the $\nu$ term is required for stability, 
as either $\mu_x$ or $\mu_1$ vanishes at criticality \footnote{%
Other derivative terms of the same order are in principle allowed, 
but since they are irrelevant and only regularize the short-wavelength limit, 
our results are independent of this isotropic choice.}.

The above equation corresponds precisely to the nonconserved 
(Model~A) dynamics of the linearized Landau free energy 
at a $(d,m)$ Lifshitz point~\cite{chaikin_b95,hornreich_jmmm80}:
\begin{equation}
	F_{d,m} = 
	\frac{1}{2} \! \int_{\br} 
	\!\Big[
		M \phi^2 
		+ c_\parallel (\nabla_\parallel \phi)^2 
		+ c_\bot (\nabla_\bot \phi)^2 
		+ D (\nabla^2 \phi)^2
	\Big] ,
\end{equation}
where $\phi$ is an $O(n)$-symmetric vector field 
and the $d$-dimensional space is decomposed into 
$(d\!-\!m)$ and $m$-dimensional subspaces 
$\br_\parallel$ and $\br_\bot$. 
At the critical Lifshitz point, one fine-tunes 
$c_\bot = 0$ and $M = 0$, 
leading to the scaling dimension
\begin{equation}
\label{eq:Lifshitz}
	\chi_\phi = 2 + \frac{m}{2} - d \, .
\end{equation}

Returning to our active fluid model, 
the critical point associated with $C_2$ 
(i.e., $\mu_1 = 0$) can be mapped onto this equilibrium framework 
by identifying $\bg_T \!\leftrightarrow\! \phi$, 
$n = d-2$, $\mu_1 \!\leftrightarrow\! c_\bot$, $m = d-1$, and $\nu = D$. 
Unlike in equilibrium, however, the $\bg_T$ modes here 
are Goldstone modes of the TT equation in the symmetry-broken phase, 
and are therefore massless ($M = 0$) by construction, 
protected by rotational symmetry without further fine-tuning. 
From Eq.~\eqref{eq:Lifshitz}, we obtain
\begin{equation}
\label{eq:C2exps}
	\chi^{C_2}_T = \frac{3 - d}{2} , \quad 
	z^{C_2} = 4 , \quad 
	\zeta^{C_2} = 2 \, .
\end{equation}

A similar correspondence holds for the critical point associated with $C_3$ 
(i.e., $\mu_x = 0$), where we identify $\mu_x \!\leftrightarrow\! c_\bot$ and $m = 1$. 
In this case,
\begin{equation}
\label{eq:C3exps}
	\chi^{C_3}_T = \frac{5 - 2d}{4} , \quad 
	z^{C_3} = 2 , \quad 
	\zeta^{C_3} = \frac{1}{2} \, .
\end{equation}

However, as discussed in Sec.~\ref{sec:C1to4}, whenever the $C_3$ instability is triggered, so is the $C_4$ instability due to the merging of two eigenvalues. In the case that $E_-(\theta=0)>E_+(\theta=0)$ (or equivalently $E_+(\theta=\pi)>E_-(\theta=\pi)$), the $C_4$ instability becomes critical at the same time. The critical scaling behaviour of $C_4$ is however different from that of $C_3$, which will be discussed in the next section.

\section{Two critical points from $C_4$}
Critical behavior associated with $C_4$ corresponds to instabilities in the 
$E_{\pm}$ eigenvalues \eqref{eq:evs}, 
which involve only the $\rho$ and $\bg_L$ fields. 
To analyze the critical scaling at the linear level, 
we first consider the equal–time correlation functions 
$C_{\rho\rho}$ and $C_{LL}$ in the ordered phase, 
away from criticality \cite{toner_pre12}:
\begin{align}
\label{eq:Crr}
C_{\rho\rho} &= 
\int_q e^{\ii \bq \cdot \br} 
D \, 
\frac{D B_1(\theta)\sin^2\!\theta }
{ { 4} q^2 \!\left[ A_2(\theta) B_1(\theta)^2 - B_2(\theta)^2 \right] } \, , \\[4pt]
\label{eq:CLL}
C_{LL} &= 
\int_q e^{\ii \bq \cdot \br} 
D \, 
\frac{
\big[\xi^2 \cos^2\!\theta + A_2(\theta)\big] B_1(\theta) 
{ -2} \xi B_2(\theta) \cos\!\theta
}{
{ 4} q^2 \!\left[ A_2(\theta) B_1(\theta)^2 - B_2(\theta)^2 \right]
} \, ,
\end{align}
where $\xi = (\alpha_1  + \lambda_1 \beta g_0^2)/(2 \beta g_0)$.

To extract the critical scaling exponents, 
we analyze how $C_{\rho\rho}$ and $C_{LL}$ diverge as $q \!\to\! 0$ 
when $C_4$ satisfies the criticality conditions~\eqref{eq:crit_cond}. 
These can, in principle, be met for any critical polar angle $\theta_c$, 
and the resulting scaling behavior depends on whether $\theta_c = 0$ 
or $\theta_c \!\neq\! 0$.  
Without loss of generality, we restrict to $\theta \in [0, \pi/2]$.

\subsection{Case $\theta_c = 0$}
Setting $\theta_c = 0$, 
the denominators in Eqs.~\eqref{eq:Crr}–\eqref{eq:CLL} 
vanish at $\theta = 0$, 
but the angular integrations remain convergent 
(at least for $d > 3$) \cite{SM}.  
We thus find
\begin{equation}
\label{eq:C4aexps}
\chi^{C_{4a}}_\rho = \chi^{C_{4a}}_L = \frac{2 - d}{2} , 
\quad z^{C_{4a}} = 4 , 
\quad \zeta^{C_{4a}} = 1 \, .
\end{equation}

Depending on which of the two eigenvalues $E_\pm$ caused the criticality, as described above, the $C_3$ critical point might coincide with the critical behaviour described here.

\subsection{Case $\theta_c \neq 0$}
When $\theta_c \!\neq\! 0$, 
the denominator in the correlation functions 
vanishes at a finite angle, 
and the angular integrals can no longer be evaluated directly.  
As in the $C_2$/$C_3$ cases, the damping coefficients must be regularized 
by including higher–order derivatives, 
implemented here by redefining 
$B_1(\theta) \!\to\! B_1(\theta) - \nu q^2$.  
Isolating the angular contributions reveals the $q$–dependence of the correlations, 
and the detailed derivation is given in \cite{SM}. 
The resulting exponents are
\begin{equation}
\label{eq:C4bexps}
\chi^{C_{4b}}_\rho = \chi^{C_{4b}}_L = \frac{3 - d}{2} , 
\quad z^{C_{4b}} = 4 , 
\quad \zeta^{C_{4b}} = 1 \, .
\end{equation}

Both critical points exhibit distinct and inherently nonequilibrium 
scaling behavior already at the mean-field level. 
It is therefore plausible that, upon inclusion of relevant nonlinearities, 
their anomalous scaling persists down to physical dimensions.

The analysis of the three new criticality criteria---$C_2$, $C_3$, and $C_4$---for the Toner–Tu model thus uncovers at least two new classes of 
nonequilibrium critical phenomena (both associated with $C_4$), 
as evidenced by their distinct sets of linear critical exponents, summarized in Table~\ref{tab}. This constitutes the second key result of this paper.

\begin{table}
\begin{tabular}{p{0.1\linewidth}|p{0.18\linewidth} |p{0.22\linewidth}| p{0.1\linewidth}| p{0.1\linewidth}| p{0.2\linewidth}}
\hline
Type& Crit.~pt. & $\chi$'s & $z$ & $\zeta$ & Pertinent\newline coefficient\\
\hline
\hline
I & $C_1$ \cite{miller_pre24b,miller_pre24} & $  \chi_{\rho}=\frac{3-d}{2}$  & $ 2$ & $ 2$ & $\kappa_1$ \\
II& $C_2$ & $\chi_T=\frac{3-d}{2}$ & $4$ & $2$ & $\mu_\perp$\\
II& $C_3$ & $\chi_T=\frac{5-2d}{4}$ & 2 & $\frac{1}{2}$ & $\mu_x$\\
II& $C_{4a}$ & $\chi_{L /\rho}=\frac{2-d}{2}$ & $4$ & $1$ & $\lambda_3$\\
II& $C_{4b}$ & $\chi_{L /\rho}=\frac{3-d}{2}$ & $4$ & $1$ &$\kappa_1$\\
\hline
\end{tabular}
\caption{{\it Comprehensive classification of critical points in the ordered Toner–Tu phase.} Each row corresponds to one of the instability criteria $C_i$, indicating the type of instability, the associated critical fields, the mean-field critical exponents $(\chi, z, \zeta)$, and the hydrodynamic coefficient whose fine-tuning drives the system to criticality (besides $\rho$). The list is comprehensive for all generic (non-multicritical) critical points---i.e., those accessible by tuning only two parameters—and includes both previously known cases ($C_1$) and the new critical points identified in this work ($C_2$, $C_3$,  $C_{4a}$, $C_{4b}$).  
The critical point $C_3$ can  only be reached simultaneously with the $C_{4a}$ criticality (but not vice versa), due to the merging of two eigenvalues as discussed in Section~\ref{sec:C1to4}.
}
\label{tab}
\end{table}

\begin{figure*}
\includegraphics[width=\linewidth]{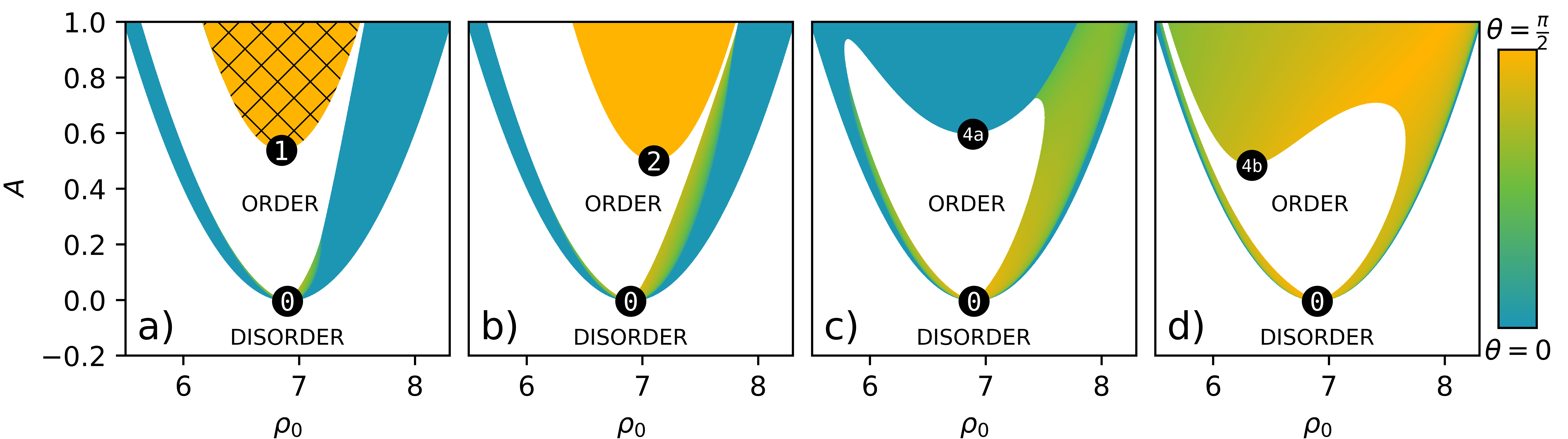}
\caption{{\it Linear instability regions and associated critical points in the ordered phase
of the Toner--Tu model.}
Colored regions indicate long-wavelength (hydrodynamic) instabilities in the $q \to 0$ limit; the color encodes the direction of the most unstable wavevector.  Hatched regions mark  {\it Type I}  instabilities. Critical points are labeled by the numbers used in the text: (a) “0” denotes the order--disorder transition associated with $C_0$ \cite{nesbitt_njp21,bertrand_prr22}; “1” marks the  critical point recently reported in \cite{miller_pre24}. The remaining points are predictions of our stability analysis: (b) $C_2$ (see Sec.~\ref{sec:C23}); (c) $C_{4a}$ with $\theta_c = 0$ based on model Eq.~\eqref{eq:kapmodel}; and (d) $C_{4b}$ with $\theta_c \neq 0$ based on model Eq.~\eqref{eq:model2}. Since $C_3$ criticality occurs necessarily in conjunction with $C_{4a}$ criticality, it is not depicted here explicitly.
}
\label{fig:PD}
\end{figure*}

\section{Realizing the critical points in hydrodynamic models}

As shown above, the ordered phase of the Toner-Tu equations houses a diverse set of distinct critical phenomena which can, in principle, be accessed by fine-tuning only two model parameters. But can these critical points also be accessed by physically sensible microscopic models? Indeed, while the Toner-Tu equations are able to generically describe dry polar active fluids at large length and time scales, typical microscopic models usually allow tuning of very few parameters and are thus only able to explore a low dimensional subspace of the full phase diagram spanned by the complete set of hydrodynamic parameters of the TT equations. 

The Vicsek model for example does not feature the critical order-disorder transition $C_0$. This doesn't mean, however, that real flocking systems cannot exhibit this continuus phase transition. Rather it is a consequence of the absence of short-range repulsions, a reasonable simplification if one is mainly interested in the dynamics of the flocking phase, which however clearly does not apply to real systems with physical particles. On the hydrodynamic level, the absence of short-range repulsion leads to $C_0$ being montonous in $\rho_0$, thus inhibiting the critical phase transition. 
Taking into account the effect of short-range repulsion and the resulting contact inhibition of locomotion \cite{romenskyy_epjb13, schnyder_scirep17} yields to a nonmonotonous $C_0$ and thus a realization of the critical order--disorder transition $C_0$ \cite{nesbitt_njp21,bertrand_prr22,agranov_njp24}.

To lowest order in the vicinity of the critical transition, this effect can be modeled on the hydrodynamic level, by expanding the function $U$, 
\begin{equation}
	\label{eq:basis}
  U = -A + a_1(\rho - \rho_{\rm ref})
  + \frac{1}{2}\Big[a_2(\rho - \rho_{\rm ref})^2 + \beta\mathbf{g}^2 \Big] ,
\end{equation}
which controls the mean velocity of the flocking phase to second order in the fields, resulting in 
\begin{equation}
  C_0 = 2\Big(- A + a_1(\rho - \rho_{\rm ref})
  + \frac{1}{2} a_2(\rho - \rho_{\rm ref})^2 \Big) ,
\end{equation}
realizing the critical phase transition at $A = 0$ and $\rho=\rho_{\rm ref}$ (see black circle labelled ``0'' in Fig.~\ref{fig:PD}).

In the following we show that the remaining critical behaviors can likewise be motivated by physically sensible mechanisms, thereby making our predictions experimentally relevant. Using the model defined by Eq.~\eqref{eq:basis} as a basis, we propose for each critical point $C_i$, a physical mechanism that would modify the relevant hydrodynamic parameters such that criticality is realized.

Adopting a convenient choice of units (time, length, mass), we set $a_1 = \frac{1}{10}, \quad a_2 = 1, \quad \beta = 1, \quad \rho_{\rm ref}=7$, $\lambda_3=-1/2$ and, unless otherwise stated, all other coefficients in the TT equations to $1$. 

We assume that the two experimentally tunable parameters are the particle density $\rho_0$ as well as the activity parameter $A$, which sets the speed of collective motion $A$ at the reference density $\rho_{\rm ref}$.

\subsection{$C_1$ criticality}
\label{sec:C1}

For completeness and coherence, we outline how $C_1$ criticality can be realized within the present framework (see also prior discussions in the literature).

Although the inverse compressibility, $\kappa_1$, is typically positive, 
Miller and Toner~\cite{miller_prl24} have shown that attractive interactions
such as chemotactic couplings can drive it negative. 
 Here we assume that flocking particles are actively attracted to each other. Hydrodynamically, an increase in activity will thus lead to a decrease in inverse compressibility $\kappa_1$ (i.e., an increase in compressibility). Further, in the absence of activity, we assume that particles have a preferred packing density $\rho_{\rm opt}$, resulting in the compressibility being maximal at this value. Both these features are reasonable for a range of natural and engineered active-fluid systems.

A convenient parametrization consistent with these assumptions is
\begin{equation}
	\label{eq:kapmodel}
  \kappa_1 \;=\; k_0 \;-\; k_A A \;+\; (\rho - \rho_{\rm opt})^2 \, .
\end{equation}

Fig.~\ref{fig:PD}(a) shows the region of linear instability, colored according to the direction of the most unstable wavevector ({\it Type I} instabilities are hatched) together with the $C_1$ critical point (black circle labeled ``1''), using the illustrative choices: $\rho_{\rm opt}=7.1$, $k_0= 1/2$ and $k_A=1$.

\subsection{$C_2$ \& $C_3$ criticality}

The $C_2$ and $C_3$ critical points belong to the Lifshitz universality class \cite{chaikin_b95,hornreich_jmmm80}, which in the general case is anisotropic with two $m$ and $d-m$ dimensional subspaces, corresponding here to the flocking direction ${\hat \bx}$ and the directions orthogonal to it ${\hat \bx_\bot}$. The instability is triggered by the effective viscosities $\mu_x$ or $\mu_\bot$ becoming negative. These control how flocking particles align their velocity with neighboring particles. A positive $\mu_{x/\bot}$ corresponds to an alignment with particles in front or behind/left and right. These alignments can have both passive and active origins, in sum giving rise to an effective viscosity. Strong enough (anisotropic) anti-alignment can thus also give rise to negative viscosities. At large densities, passive viscous effects will however still dominate over active anti-alignment, giving rise to stabilizing $\nu q^4$ at higher order in wave number \cite{SM}. Such an effect has indeed been used to described active turbulent flow in bacterial colonies \cite{wensink_pnas12}. 

To realize the $C_2$ universality class we thus adopt a model, where activity lowers isotropic particle viscosity. To realize a nonmonotonous density dependence of the viscosity we also assume that the interactions between particles is topological, i.e., the preference of particles to align or anti-align is not dependent on the physical distance between the two, but rather on the degree of neighborhood. On the hydrodynamic level this implies that viscosity must increase towards lower densities. At the same time for higher densities, passive viscous effects are expected to dominated over active alignment. This naturally gives rise to density optimum $\rho_{\rm LP}$ where viscosity is minimal, such that around this minimum, we can assume the velocity has the form,
\beq
\mu_{1} = m_0 - m_A A + \big(\rho - \rho_{\rm LP}\big)^2 \, .
\eeq
Fig.~\ref{fig:PD}(b) shows the  the phase diagram corresponding to this model with the $C_{2}$ critical point (black circle labeled ``2''), using the illustrative choices $\rho_{\rm LP}=7.1$, $m_0 = 1/2$, and $m_A=1$.

Since the  the $C_3$ critical point, cannot be independently reached from $C_{4a}$, we do not explicitly construct a model for it. Nevertheless, the considerations above can straightforwardly be made directionally dependent and applied to $\mu_3$ to realize the $C_3$ critical point.

\subsection{$C_{4a}$ criticality}

To realize $C_{4a}$ criticality, we revisit the variable–compressibility model of Sec.~\ref{sec:C1} and select a parameter regime in which negative values of $\kappa_1$ activate \emph{only} the $C_4$ instability (and not $C_1$). Concretely, we (i) choose a relatively large baseline $k_0$, and (ii) set the longitudinal density–coupling to $\nu_1 = -3$. Physically, the latter choice corresponds to particles being more strongly attracted to neighbors directly ahead of or behind them than to those located laterally.

We parameterize the compressibility as in Eq.~\eqref{eq:kapmodel}, however with parameter choices
$k_0 = 4$, $k_A = 2$ and $\rho_{\rm opt}=6$.  With this particular model, $C_{4a}$ criticality is reached without the $C_3$ instability being triggered.

The resulting phase diagram [Fig.~\ref{fig:PD}(c)] exhibits an extended region of instability whose most unstable wavevector aligns with the flocking direction, as expected for the $C_{4a}$ critical point (black circle labeled ``4a''). 

\subsection{$C_{4b}$ criticality}

To realize the $C_{4b}$ criticality, we exploit the fact that the critical angle $\theta_c$ depends sensitively on the coupling $\lambda_3$~\cite{SM}, which controls how the momentum density responds to gradients of the local speed. We therefore expand $\lambda_3$ in both the control parameter $A$ and the mean density $\rho_0$:
\begin{equation}
\label{eq:model2}
    \lambda_3 \;=\;
    l_0 \;+\; l_A\,A \;+\; l_1\,(\rho_0 - \rho_{\rm ref})
    \;+\; \frac{l_2}{2}\,(\rho_0 - \rho_{\rm ref})^2 \, ,
\end{equation}
with $l_0 = \tfrac{3}{2}$, $l_1 = 2$, $l_2 = -2$, and $l_A = 5$.
Physically, the choice in Eq.~\eqref{eq:model2} implies that particles generally bias motion toward regions of higher speed, except within a narrow density window where this preference reverses—thereby selecting a finite critical angle $\theta_c \neq 0$ characteristic of the $C_{4b}$ instability.

To suppress competing instabilities, we additionally fix $\kappa_1 = 20$ and $\nu = 1$. With these choices, the model realizes the $C_{4b}$ critical point, as shown in Fig.~\ref{fig:PD}(d) (black circle labeled ``4b'').

\vspace{.2in}

The hydrodynamic constructions above show that each critical phenomenon identified here can arise within distinct regions of the Toner-Tu parameter space. Although the parametrizations may appear ad hoc at first glance, we have grounded them in plausible microscopic mechanisms, specifying how speed-, density-, and direction-dependent responses of active agents generate the required couplings, constituting our third key result. In line with Feynman's dictum---``what is not forbidden is compulsory''---we therefore expect these scenarios to be realized across a broad range of natural and engineered active systems.

\section{Summary and outlook}
We have performed a comprehensive linear stability analysis of the Toner–Tu model for polar active fluids in the ordered phase and derived, in fully analytical form, the corresponding instability criteria. Our treatment provides a complete classification of all hydrodynamic instabilities into two generic types, distinguished by how their growth rates scale with the wavevector magnitude. By applying the criticality conditions, we showed that fine-tuning only two control parameters is sufficient to reach criticality for all instability types.

Among the resulting critical points, four have not been previously reported for the Toner–Tu model; notably, two of these (associated with the $C_4$ condition) exhibit nonequilibrium critical behavior that already departs from known universality classes at the linear level. Another central contribution of this work is that we explicitly constructed hydrodynamic models, rooted in concrete physical mechanisms, that realize each of these new critical points, demonstrating that the full diversity of predicted critical phenomena is not merely an abstract possibility of the stability analysis, but can arise robustly within physically motivated active-fluid theories. These realizations also clarify the physical mechanisms—such as sign changes in effective inverse compressibility, viscosity or speed-response couplings—that underpin the different classes of criticality.

An important next step is to identify microscopic (agent-based or kinetic) models that give rise to these hydrodynamic realizations and to investigate the associated universality classes using renormalization group (RG) methods. The rapid expansion in known nonequilibrium universality classes over the past decade has been driven not by conceptual breakthroughs in RG, but by understanding where to look. We hope that the analytical framework and explicit realizations provided here will serve as a roadmap for uncovering further nonequilibrium universality classes in active matter.

\bibliography{references}

\end{document}


\title{Supplemental material for ``Diversity of critical phenomena in the ordered phase of polar active fluids"}
\author{Patrick Jentsch}
	\email{patrick.jentsch@embl.de}
	\affiliation{Cell Biology and Biophysics Unit, European Molecular Biology Laboratory Heidelberg, Meyerhofstrasse 1, 69117 Heidelberg, Germany}
	\author{Chiu Fan Lee}
	\email{c.lee@imperial.ac.uk}
	\affiliation{Department of Bioengineering, Imperial College London, South Kensington Campus, London SW7 2AZ, U.K.}
\date{\today}

	\begin{abstract}
	\end{abstract}
	
\maketitle

\section{The Toner-Tu Hydrodynamic Equation of Motion}

We use the general form of the Toner-Tu equations \cite{toner_pre12}, containing all terms allowed by translational, rotational and chiral invariance, as well as mass conservation. The main difference with \cite{toner_pre12} is that we use the momentum density $\bg$ as the order parameter for collective motion, rather the velocity $\bv=\bg/\rho$. As a consequence, the continuity equation for the density $\rho$ is linear in the fields,
\begin{equation}
	\label{eq:cont}
	\partial_t \rho + \nabla\cdot \bg = 0 \ .
\end{equation}
The equations of motion (EOM) for the momentum density is,
\begin{align}
	\label{eq:tt}
	\nonumber
	\partial_t g_i &+\lambda_1 g_j \nabla_j g_i + \lambda_2 g_i \nabla_j g_j+\lambda_3 g_j \nabla_i g_j =-Ug_i-\kappa_1 \nabla_i \rho  -  \nu_1 g_i g_j \nabla_j \rho\\
	&     +\mu_1 \nabla^2 g_i+ \mu_2 \nabla_i \nabla_j g_j + \mu_3 g_j g_k\nabla_j\nabla_k g_i + \mu_4 g_i g_j \nabla^2 g_j+\mu_5 g_i g_j \nabla_j \nabla_k g_k+\mu_6 g_j g_k \nabla_k \nabla_i g_j +\mathrm{h.o.t.}+\bff  \ ,
	\end{align}
where all couplings, i.e., $\lambda_1$, $\lambda_2$, $\lambda_3$, $U$, $\kappa_1$, $\nu_1$, $\mu_1$, $\mu_2$, $\mu_3$, $\mu_4$, $\mu_5$ and $\mu_6$, depend in general on $\rho$ and $\phi=|\bg|^2/2$. ``h.o.t.'' denotes higher order terms, i.e., terms of higher order in derivatives. Further, some terms of order $\mathcal O(\nabla^2 g^3)$, where the spatial derivatives are acting on different instances of $\bg$, namely,
\begin{align}
	\nonumber
	g_j \nabla_i g_j \nabla_k g_k \sep g_j \nabla_i g_k \nabla_j g_k \sep g_j \nabla_i g_k \nabla_k g_j \ , \\
	\nonumber
	g_j \nabla_j g_i \nabla_k g_k \sep g_j \nabla_k g_i \nabla_j g_k  \sep	g_j \nabla_k g_i \nabla_k g_j \ , \\
	g_i \nabla_j g_k \nabla_j g_k \sep g_i \nabla_j g_k \nabla_k g_j \sep g_i \nabla_j g_j \nabla_k g_k \ ,
\end{align}
have been neglected, since they do not contribute to the linear stability analysis. Finally, $\bff$ is a Gaussian noise term with vanishing mean and statistics,
\begin{equation}
\langle f_i(t,\br) f_j(t^\prime,\br^\prime)  \rangle = 2D\delta_{ij}\delta^{d+1}(t-t^\prime,\br-\br^\prime) \ ,
\end{equation}
where $\delta^{d+1}$ is the $(d+1)$-dimensional Dirac delta function. 

Since we are interested in instabilities of the ordered state, we assume that the system is in its ordered phase, i.e., given the average density $\rho_0$, there exists a $\phi_0$ such that $U(\rho_0,\phi_0)=0$, $U(\rho_0,\phi)$ is strictly negative for $\phi <\phi_0$ and strictly positive for $\phi>\phi_0$. The set $(\rho_0,\bg_0)$ with $|\bg_0|^2 = g_0^2= 2 \Phi_0$ is thus a solution to the mean-field EOM.

\section{Linear Stability Analysis}
\label{sec:linstab}

We now linearly expand the EOM around this homogeneous solution. Without loss of generality, the background momentum density is chosen to point into the $x$-direction, i.e., $\bg_0 = g_0 \hat \bx$, and,
\begin{equation}
\rho(t,\br) = \rho_0 + \delta \rho(t,\br) \ , \ \ \ \ \bg(t,\br) = \bg_0 + \bg_x(t,\br) + \bg_\bot(t,\br)  \ , \ \ \ \ \phi(t,\br) = \phi_0 +\bg_0\cdot \bg_x(t,\br)
\end{equation}
where one distinguishes the fluctuations $\delta\bg=\bg-\bg_0$ between fluctuations along the flocking direction $\bg_x = g_x \hat\bx$ and perpendicular to it $\bg_\bot$. The linearized EOM become,
\begin{equation}
\label{eq:linear_eom_1}
\partial_t\delta\rho+\nabla_\bot\cdot\bg_\bot + \partial_x  g_x = 0 \ ,
\end{equation}
\begin{align}
	\label{eq:linear_eom_2}
	\nonumber
	\partial_t \bg_x &+(\lambda_1+\lambda_2 +\lambda_3)g_0 \partial_x \bg_x + \lambda_2 \bg_0 \nabla_\bot \cdot \bg_\bot=-\beta g_0^2 \bg_x - \alpha_1 \bg_0 \delta\rho - \kappa_1 \hat\bx \partial_x \delta\rho - \nu_1 \bg_0 g_0 \partial_x \delta\rho \\
	&     +\Big[(\mu_1 + \mu_4 g_0^2) \nabla^2_\bot+(\mu_1+\mu_2 + \mu_3 g_0^2+\mu_4 g_0^2+\mu_5 g_0^2+\mu_6 g_0^2)\partial_x^2\Big] \bg_x+ (\mu_2 +\mu_5 g_0^2) \hat\bx \partial_x \nabla_\bot \cdot \bg_\bot   +\bff \ ,
\end{align}
and
\begin{align}
	\label{eq:linear_eom_3}
	\nonumber
	\partial_t \bg_\bot &+\lambda_1 g_0 \partial_x \bg_\bot +\lambda_3 g_0 \nabla_\bot g_x = \\
	&- \kappa_1 \nabla_\bot \delta\rho +\Big[\mu_1 \nabla^2_\bot+(\mu_1+\mu_3g_0^2)\partial_x^2\Big] \bg_\bot + \mu_2 \nabla_\bot \nabla_\bot \cdot \bg_\bot+(\mu_2  +\mu_6 g_0^2)\nabla_\bot \partial_x\bg_x +\bff  \ ,
	\end{align}
where,
\begin{equation}
\beta =  U^{(1,0)}(\phi_0,\rho_0) \ ,   \ \ \alpha_1 =  U^{(0,1)}(\phi_0,\rho_0) \ , 
\end{equation}
and all other couplings are also evaluated at $\phi=\phi_0$ and $\rho=\rho_0$. The linear EOM \eqref{eq:linear_eom_1}-\eqref{eq:linear_eom_3} are transferred into Laplace-Fourier transformed space (Laplace in time and Fourier in space, i.e., $\delta \rho(t, \br) = \delta \rho \ee^{st-\ii \bq \cdot \br}$, etc). Then the perpendicular field $\bg_\bot$ is further split into its longitudinal and transverse components,
\begin{equation}
\bg_L = \hat\bq_\bot (\hat\bq_\bot \cdot \bg_\bot) \ , \ \ \ \bg_T =\bg_\bot - \bg_L \ ,
\end{equation}
i.e., the component parallel and perpendicular to the wavevector $\bq_\bot=\bq-q_x\hat\bx$, where $q_x=\bq\cdot \hat\bx$, $q = |\bq|$, and $q_\bot = |\bq_\bot|$. Hats represent normalized vectors.
The linearized EOM \eqref{eq:linear_eom_1}-\eqref{eq:linear_eom_3} thus becomes an Eigenvalue problem,
\begin{equation}
\label{eq:eigenvalue}
M \cdot \begin{pmatrix}
	\delta \rho \\ \delta\bg
\end{pmatrix}
=
s\begin{pmatrix}
	\delta \rho \\ \delta\bg
\end{pmatrix}
\ ,
\end{equation}
where the matrix $M$ can be written as a block matrix,
\begin{equation}
M =
\begin{pmatrix}
N & { 0} \\
{ 0} &-\left(\ii \lambda_1 g_0 q_x + \mu_x q_x^2+\mu_1 q_\bot^2 \right) \bI_{d-2}
\end{pmatrix} \  ,
\end{equation}
defined in terms of  $(d-2)$-dimensional unit matrix $\bI_{d-2}$ and the fully coupled matrix $N$,
\begin{equation}
N=
\begin{pmatrix}
0 & -\ii q_x & -\ii q_\bot  \\
-\alpha_1 g_0 - \ii \kappa_x q_x & -\ii \lambda_{\rm tot} g_0 q_x -\beta g_0^2 -\mu^{xx}_x q_x^2-\mu^{xx}_\perp q_\bot^2 & -\ii \lambda_2 g_0 q_\bot-\mu^{xL} q_x q_\bot \\
-\ii \kappa_1 q_\bot & -\ii \lambda_3 g_0 q_\bot - \mu^{Lx} q_xq_\bot & -\ii \lambda_1 g_0 q_x - \mu_x q_x^2-\mu^{LL}_\bot q_\bot^2
\end{pmatrix} \ ,
\end{equation}
with,
\begin{equation}
\kappa_x = \kappa_1 +\nu_1 g_0^2 \ , \ \ \ \lambda_{\rm tot} = \lambda_1+\lambda_2+\lambda_3 \ ,  
\end{equation}
and
\begin{align}
	\mu^{xx}_\perp &= \mu_1 +\mu_4 g_0^2 \sep& \mu^{xx}_x &= \mu_1+\mu_2 + \mu_3 g_0^2+\mu_4 g_0^2+\mu_5 g_0^2+\mu_6 g_0^2 &\sep \mu^{xL} &= \mu_2 +\mu_5 g_0^2 \ , \\
	\mu^{LL}_\perp &= \mu_1 +\mu_2 \sep& \mu_x &= \mu_1 + \mu_3 g_0^2 &\sep \mu^{Lx} &= \mu_2 +\mu_6 g_0^2 \ .
\end{align}

Since $N$ is a fully coupled $3 \times 3$ matrix, the expressions for its general eigenvalues are lengthy but can be expressed analytically  (see {\it supplemental\textunderscore notebook.nb}). For the phase diagram however, only the large-scale instabilities are necessary, which corresponds to the small $q$-limit. In this limit, the eigenvalues of $M$ become,
\begin{align}
	\label{eq:eigenvalues_stab_ordered}
	s &\xrightarrow{q\rightarrow0}  \left\{\begin{array}{ll}
	E_0 &=-\beta g_0^2 \\
	E_{\pm} &=\ii q A_1(\theta)\pm \ii q \sqrt{A_2(\theta)} - q^2 B_1(\theta)\pm  q^2 \frac{B_2(\theta)}{\sqrt{A_2(\theta)}} \\
	E_T &= -\ii \lambda_1 g_0 q_x-\mu_x q_x^2 - \mu_1 q_\bot^2
	\end{array}
	\right. \ ,
	\end{align}
to leading order in $q$ for the real part of each eigenvalue respectively. Further, we have defined the functions,
\begin{align}
	\label{eq:A1}
	A_1(\theta) &= \cos(\theta)\frac{\alpha_1 -\beta g_0^2 \lambda_1 }{2\beta g_0} \ , \\
	\label{eq:A2}
	A_2(\theta) &= \frac{1}{4 \beta^2 g_0^2} \Bigg[\cos^2(\theta)\Big(\alpha_1 + \lambda_1 \beta g_0^2\Big)^2+4 \sin^2(\theta)\Big(\beta  \kappa_1-\alpha_1    \lambda_3 \Big)\beta g_0^2  \Bigg] \ ,
\end{align}
and,
\begin{align}
\nonumber
B_1(\theta) = \frac{1}{2\beta^3 g_0^4} \Bigg[ &\cos^2(\theta) \Big(\beta^2 g_0^2 (\kappa_x+\beta g_0^2 \mu_x)-\alpha_1 (\alpha_1 + \lambda_{\rm tot} \beta g_0^2) \Big) \\ 
&+ \sin^2(\theta)\beta g_0^2\Big(\alpha_1 \lambda_3 +\beta g_0^2(\lambda_2\lambda_3 +\beta  \mu_\bot^{LL})\Big)\Bigg] \\
\nonumber
B_2(\theta) = \frac{\cos(\theta)}{4 \beta^4 g_0^5} \Bigg[& \cos^2(\theta)  (\alpha_1+\lambda_1\beta g_0^2)\Big(\beta^2 g_0^2 (\beta g_0^2 \mu_x-\kappa_x)+\alpha_1 (\alpha_1 + \lambda_{\rm tot} \beta g_0^2 )\Big) \\
\nonumber
&+\sin^2(\theta)\beta g_0^2\Big(\alpha_1\beta (2\kappa_1+(\lambda_1 -\lambda_2-2\lambda_{\rm tot}) \lambda_3 g_0^2-2\beta g_0^2 \mu^{Lx}+\beta g_0^2 \mu_\bot^{LL})-3\alpha_1^2 \lambda_3 \Big) \\
&+\sin^2(\theta) \beta^3 g_0^4 \Big(2\kappa_1\lambda_2 +2\kappa_x\lambda_3 +\lambda_1\lambda_2\lambda_3 g_0^2+\beta\lambda_1 g_0^2\mu_\bot^{LL}\Big)
\Bigg] \ ,
\end{align}
and have expressed the wavevector in polar coordinates,  $q_x = q \cos(\theta)$ and $q_\bot = q \sin(\theta)$.

\subsection{Type I instability}

Going through the eigenvalues \eqref{eq:eigenvalues_stab_ordered} order by order, one finds that the first eigenvalue $E_0$ is always negative in the ordered phase. At linear order in $q_x$ and $q_\bot$, the eigenvalues $E_\pm$ and $E_T$ are purely imaginary if $A_2(\theta)>0$ and thus not destabalizing.

However, if,
\begin{equation}
	A_2(\theta)<0 \ ,
\end{equation}
for at least some values of $\theta$, $E_-$ will develop a positive real part at linear order in $q$ leading to an instability. We term this a {\it Type I} instability, due to the linear scaling in $\bq$. Generically, since the rotational symmetry is broken in the ordered phase, not all wave-vectors are equally unstable, and one can further classify the instability according to the direction of the most unstable mode. Since $A_2$ is invariant under $\theta \rightarrow -\theta$ and $\theta \rightarrow \pi-\theta$, we can, without loss of generality, restrict to the case $\theta\in[0,\pi/2]$. Due to the positive definitiveness of the first term of $A_2$, \eqref{eq:A2}, the most unstable direction for a {\it Type I} instability is always $\theta=\pi/2$. Therefore, the instability condition can be simplified to,
\begin{equation}
	C_1\equiv -\beta A_2(\pi/2) = \alpha_1\lambda_3-\beta\kappa_1 > 0 \ ,
\end{equation}
which is the instability condition derived in \cite{miller_pre24b}. Since $\lambda_1$ and $g_0$ are real numbers in the ordered phase, the eigenvalue $E_T$ cannot have a {\it Type I} instability.

\subsection{Type II instability}

We now turn to the case where $A_2>0$ for all $\theta$, and explore instabilities at the next order in wavenumber, i.e., at order $q^2$, which we term {\it Type II} instability. The most unstable directions of the eigenvalue $E_T$ are  clearly the values $\theta=0$ and $\theta=\pi/2$, with the corresponding instabilities,
\begin{equation}
	C_2\equiv - \mu_1 >  0 \sep \text{and} \sep C_3\equiv -\mu_x > 0 \ .
\end{equation}
For the eigenvalue $E_\pm$ the situation is much richer. Due to the more complicated angular dependence, the ordered phase can become unstable if,
\begin{equation}
	\label{eq:instab_type2}
	-B_1(\theta) + \Bigg|\frac{B_2(\theta)}{\sqrt{A_2(\theta)}} \Bigg| > 0 \ ,
\end{equation} 
for any value of $\theta$. The absolute value comes from the fact that the whole system becomes unstable if either of the two eigenvalues $E_\pm$ is unstable.  Again, this condition, is symmetric under reflections of the $\hat \bx$ and $\hat \bq_\bot$ axis, i.e., under $\theta\rightarrow-\theta$ and $\theta\rightarrow\pi-\theta$, such that the instability condition must only be checked for  $\theta\in[0,\pi/2]$. We can therefore write the final instability condition as a supremum over the quarter circle,
\beq
C_4  \equiv \sup_{\theta\in[0, \pi/2]} \Bigg[ - B_1(\theta)   +\Bigg|\frac{B_{2}(\theta)}{\sqrt{|A_2(\theta)|}} \Bigg|\Bigg]>0 \ .
\eeq
Generically, the instability thus always occurs along two axis, reminiscent of the cross-sea phase observed in the Vicsek model \cite{kuersten_prl20}, except for the special values $\theta = 0$ and $\theta = \pi$.

As we will see in the next section, for the scaling behaviour near the critical points, the instabilities in the two special cases $\theta = 0$ and $\theta = \pi$ behaves differently from a generic value of $\theta$, on the level of the linear correlation functions. We therefore now investigate the instability condition \eqref{eq:instab_type2} in these special cases. First, for $\theta = 0$, we find that Eq.~\eqref{eq:instab_type2} reduces to,
\begin{equation}
	\label{eq:stabC}
	\alpha_1^2+\alpha_1 \lambda_{\rm tot} \beta g_0^2-\beta^2 g_0^2 \kappa_x > 0  \ .
\end{equation}
For the $\theta=0$ direction to be also the most unstable direction, the left-hand-side of Eq.~\eqref{eq:instab_type2} must also have a maximum at $\theta=0$. Inspecting the equation more closely, i.e., expanding it around $\theta=0$, we see that the first-order series coefficient in $\theta$ vanishes generically at this point, such that the instability condition always has an extremum at $\theta=0$. As seen in the main text, one can readily find coefficients, where this extremum becomes a global maximum, such that the associated critical point can easily be found without any additional fine-tuning.

For the case $\theta=\pi/2$ the situation is different. Evaluating Eq.~\eqref{eq:instab_type2} at $\theta=\pi/2$ yields,
\begin{equation}
	\label{eq:stabB}
	\alpha_1\lambda_3+\beta \lambda_2 \lambda_3 g_0^2+\beta^2 g_0^2 \mu_\bot^L <0 \ .
\end{equation}
However, since $B_2(\theta)$ is proportional to $\cos(\theta)$, Eq.~\eqref{eq:instab_type2} only has an extremum at that value of $\theta$ if,
\begin{equation}
	\alpha_1\beta (2\kappa_1+(\lambda_1 -\lambda_2-2\lambda_{\rm tot}) \lambda_3 g_0^2-2\beta g_0^2 \mu^{Lx}+\beta g_0^2 \mu_\bot^{LL})-3\alpha_1^2 \lambda_3  +\beta^2 g_0 \Big(2\kappa_1\lambda_2 g_0+2\kappa_x\lambda_3 g_0+\lambda_1\lambda_2\lambda_3 g_0^3+\beta\lambda_1 g_0^3\mu_\bot^{LL}\Big) = 0 \ ,
\end{equation}
i.e., to realize the corresponding critical point would require additional fine-tuning to satisfy this additional constraint. We leave this critical point therefore for future research.

\section{Critical Points}
\label{sec:critical_points}

As explained in the MT, a continuous phase transition from the stable to the unstable regime, i.e., a critical point, can only be realized if the instability criterion $C_i$ governing that phase transition is stable against local fluctuations, i.e., not only must,
\begin{equation}
	\label{eq:pre_crit_1}
	C_i(\rho_0) = 0 \, ,
\end{equation}
but also,
\begin{equation}
	\label{eq:pre_crit_2}
	C_i(\rho_0\pm \delta \rho) < 0 \, .
\end{equation}
With these conditions, fluctuations in $\bg$ are implicitly already taken into account, since all parameters are invariant under transverse fluctuations $\bg_\bot$, and fluctuations of $\bg_x$ are coupled to density fluctuations $\delta\rho$ on fast timescales, which is already taken into account by the dependence of $g_0$ on $\rho_0$. For infinitesimal $\delta\rho$, Eqns.~\eqref{eq:pre_crit_1} and \eqref{eq:pre_crit_2} are equivalent to,

\begin{equation}
	C_i(\rho_0) = 0 \sep \text{and} \sep \frac{\pp}{\pp \rho_0} C_i(\rho_0) = 0 \ ,
\end{equation}
which can be applied to all of the instability conditions given above, giving rise to a total of 5 different kinds of critical points in the ordered phase of dry polar active fluids, as we will see below.

In the following, we analyze the scaling behaviour of the critical points at the linear level.

\subsection{Critical Transverse Goldstone instability}

We first focus on the instabilities coming from the transverse Goldstone modes $E_T$ given by,
\begin{equation}
	E_T = -\ii \lambda_1 g_0 q_x - \mu_x q_x^2 - \mu_\bot^T q_\bot^2 \ ,
\end{equation}
with the two instabilities,
\begin{equation}
	C_2 = - \mu_1 > 0 \sep \text{and} \sep C_3 = -\mu_x > 0 , 
\end{equation}
which become critical when,
\begin{equation}
	\mu_1=0 \sep  \text{and} \sep \frac{\partial}{\partial{\rho_0}}\mu_1 =0 \ ,
\end{equation}
or, 
\begin{equation}
	\mu_x=0 \sep  \text{and} \sep \frac{\partial}{\partial{\rho_0}}\mu_x =0 \ ,
\end{equation}
resepectively. The rest of the instabilities are considered not to be fulfilled, such that the density and longitudinal modes have the standard linear scaling dimensions of the Toner-Tu phase,
\begin{equation}
	\chi_\rho = \chi_L = \frac{2-d}{2} \ .
\end{equation}
As described in the main text, this assumption is technically not accurate at the $C_3$ critical point, which becomes critical at the same time as the $C_{4a}$ critical point. Nevertheless, as we will see below, the scaling dimensions $\chi_\rho$ and $\chi_L$ are the same as in the TT phase.

To stabilize the theory at small scales, a fourth order derivative term, $\nu$ needs to be added to the linear theory, which for simplicity can be taken isotropic without loss of generality. After a coordinate transformation to the co-moving frame, $\br \rightarrow \br + \lambda g_0 t$, the linear EOM for $\bg_T$ thus reads,
\begin{equation}
	\partial_t \bg_T = \mu_x \partial_x^2 \bg_T + \mu_1 \nabla_\bot^2 \bg_T - \nu  \nabla^4 \bg_T \ .
\end{equation}

This EOM has precisely the form which one would get from the linearized Landau free energy of a $(d,m)$-Lifshitz point \cite{chaikin_b95},
\begin{equation}
	F_{d,m} = \frac{1}{2} \int_\br \Bigg[M \phi^2 + c_\parallel (\nabla_\parallel \phi)^2 + c_\bot(\nabla_\bot \phi)^2 +D (\nabla^2\phi)^2\Bigg] \ ,
\end{equation}
where $\phi$ is an $O(n)$ symmetric vector field and the $d$-dimensional space is partitioned into $d-m$ and $m$-dimensional subspaces $\br_\parallel$ and $\br_\bot$. At the critical point, one takes $c_\bot=0$. We can thus identify $\bg_T$ with $\phi$ for $n=d-2$  for both the $C_2$ and $C_3$ critical point. However, since the $\bg_T$ modes are the Goldstone modes of the Toner-Tu equation in the symmetry broken phase, they are massless modes, i.e., $M=0$ is protected by symmetry. Fluctuations of $\bg_T$ can only shift the mass of the $\bg_x$-mode. 
In the case of $\mu_1=0$, we can thus identify, $\mu_1 = c_\bot$ and $m=d-1$, and in the case of $\mu_x=0$, $\mu_x = c_\bot$ and $m=1$. 

For a $(d,m)$-Lifshitz point, the scaling dimension of $\phi$ is,
\begin{equation}
	\chi_\phi = 2+\frac{m}{2}-d \ .
\end{equation}
Assuming a rescaling of the transverse direction, $\br_\bot\rightarrow \br_\bot b$, for the critical point at $\mu_1=0$, this implies, 
\begin{equation}
	\chi_T = \frac{3-d}{2} \sep z = 4 \sep \zeta = 2 \ ,
\end{equation}
and similarly for the critical point at $\mu_x=0$, 
\begin{equation}
	\chi_T = \frac{5-2d}{4}  \sep z = 2 \sep \zeta = \frac{1}{2} \ .
\end{equation}
\subsection{Critical Behaviour of the sound-modes}

The critical behaviour of the {\it Type I} instability was discussed in \cite{miller_pre24}. We therefore focus on the {\it Type II} instability, based on Eq.~\eqref{eq:instab_type2}. Here, the $\delta \rho$ and $\delta \bg_L$ modes are coupled to one another but decoupled from $\bg_T$ linearly. At critical points related to Eq.~\eqref{eq:instab_type2}, $\bg_T$ therefore takes on the scaling behaviour 
\begin{equation}
	\chi_T = \frac{2-d}{2} \ ,
\end{equation}
and the scaling dimensions of the remaining two fields needs to be determined.

Away from the critical points, the equal-time density correlation function is (see {\it supplemental\textunderscore notebook.nb} and compare to \cite{toner_pre12}),
\begin{equation}
	C_{\rho\rho} = \int_q e^{\ii \bq \cdot \br}\frac{D B_1(\theta)\sin^2(\theta)}{4q^2(A_2(\theta)B_1(\theta)^2-B_2(\theta)^2)} \ ,
\end{equation}
and the longitudinal correlation function,
\begin{equation}
	C_{LL} = \int_q e^{\ii \bq \cdot \br} \frac{D\Big[\big(\xi^2 \cos^2(\theta)+A_2(\theta)\big)B_1(\theta)-2\xi\cos(\theta)B_2(\theta)\Big]}{4q^2(A_2(\theta)B_1(\theta)^2-B_2(\theta)^2)} \ ,
\end{equation}
where we have defined,
\begin{equation}
	\xi =\frac{\alpha_1+\lambda_1\beta g_0^2}{2\beta g_0} \ .
\end{equation}
Similarly to the previous section, to stabilize the instabilities occurring at the other two critical points, a fourth order derivative term needs to be added to the Toner-Tu equations. Again, to simplify the discussion, we assume that this term is isotropic without loss of generality. This term effectively shifts $B_1(\theta)\rightarrow B_1(\theta) +\nu q^2$. 

Generically, since we are considering the critical point, i.e., at the onset of instability, there will be be a single angular value $\theta = \theta_c$ where,
\begin{equation}
	-B_1(\theta_c) + \Bigg|\frac{B_2(\theta_c)}{\sqrt{A_2(\theta_c)}} \Bigg| = 0 \ . 
\end{equation}
Interestingly, the scaling behaviour is different, for the cases $\theta_c = 0$ and $\theta_c\neq 0$, corresponding to the two different critical points $C_{4a}$ and $C_{4b}$ of the main text.

\subsubsection{Critical point $C_{4b}$}

We first consider critical point $C_{4b}$, where generically $\theta_c \neq 0$. This implies in particular, that in the general case, without any further fine-tuning, all angle dependent functions, (in particular also $\sin \theta_c$ and $\cos \theta_c$) are nonvanishing when evaluated at $\theta= \theta_c$. The numerator of the integrands of the correlation functions is thus generically nonzero, such that the integral is dominated around pole at $\theta =\theta_c$ in the limit of $q\rightarrow 0$.

To take this limit, we can thus expand the $\theta$ integral around this pole, i.e., the denominator beocomes,
\begin{equation}
	A_2(\theta)\big(B_1(\theta)+\nu q^2\big)^2-B_2(\theta)^2 = 2 A_2(\theta_c) B_1(\theta_c) \nu q^2 + F^{\prime\prime}(\theta_c)(\theta-\theta_c)^2+ \mathcal O(q^2,(\theta-\theta_c)^3)
\end{equation}
where,
\begin{equation}
	F(\theta)= A_2(\theta)B_1(\theta)^2-B_2(\theta)^2 \ ,
\end{equation}
is simply the unregularized denominator. All other appearances of $\theta$ are approximated to first order, i.e., $\theta=\theta_c$, such that in spherical coordinates we find that the density correlation function is,
\begin{align}
	\nonumber
	C_{\rho\rho} &= \frac{\Gamma\Big(\frac{d}{2}\Big)S_d}{\pi\Gamma\Big(\frac{d-2}{2}\Big)(2\pi)^d}  \int_0^\infty \dd q\, q^{d-1} \int_0^\pi \dd \phi \sin(\phi)^{d-3} \int_0^\pi \dd \theta \sin(\theta)^{d-2}  e^{\ii \bq\cdot\br} \frac{D B_1(\theta)\sin^2(\theta)}{2q^2(A_2(\theta)\big(B_1(\theta)+\nu q^2\big)^2-B_2(\theta)^2)}\ , \\
	\nonumber
	&\approx \frac{\Gamma\Big(\frac{d}{2}\Big)S_d}{\pi\Gamma\Big(\frac{d-2}{2}\Big)(2\pi)^d}  \int_0^\infty \dd q\, q^{d-1} \int_0^\pi \dd \phi \sin(\phi)^{d-3} 2\int_0^{\frac{\pi}{2}} \dd \theta \sin(\theta_c)^{d-2}  e^{\ii (q x \cos \theta_c+q r_\bot \sin\theta_c \cos\phi)} \\
	&\hspace{5cm}\times \frac{D B_1(\theta_c)\sin^2(\theta_c)}{2q^2(2 A_2(\theta_c) B_1(\theta_c) \nu q^2 + F^{\prime\prime}(\theta_c)(\theta-\theta_c)^2)}\ , 
\end{align}
where we have also restricted the $\theta$ integral to the quarter circle and added a symmetry factor of $2$. We have also defined $S_d$ as the surface area of the $d$-dimensional unit sphere,
\begin{equation}
	S_d = \frac{2\pi^{d/2}}{\Gamma\Big(\frac{d}{2}\Big)}\ .
\end{equation}
In the case $\theta_c=\pi/2$, this step should be skipped, since there is only one pole in the quarter circle. This will only remove the factor $2$ from the final result and not impact the scaling behaviour. Finally, we apply a change of variables,
\begin{equation}
	\theta = m\, q\, y + \theta_c \ ,
\end{equation}
with 
\begin{equation}
	m = \sqrt{\frac{2A_2(\theta_c)B_1(\theta_c)\nu}{F^{\prime\prime}(\theta_c)}} \ ,
\end{equation}
where we have taken the positive root without loss of generality. The other root is outside the integral region. With this change of variables, the integral boundaries, get modified,
\begin{equation}
	[0,\pi]\rightarrow\Bigg[\frac{-\theta_c}{m q},\frac{\pi-2\theta_c}{2 m q}\Bigg] \ ,
\end{equation}
which, since $\theta_c$ lies within the interval $[0,\pi/2]$, in the limit of $q\rightarrow0$ become infinite boundaries. In the hydrodynamic limit, this results in,
\begin{align}
	\nonumber
	C_{\rho\rho} &=\frac{\Gamma\Big(\frac{d}{2}\Big)S_d}{\pi\Gamma\Big(\frac{d-2}{2}\Big)(2\pi)^d}  \int_0^\infty \dd q\, q^{d-1} \int_0^\pi \dd \phi \sin(\phi)^{d-3} 2\int_0^{\frac{\pi}{2}} \dd \theta \sin(\theta_c)^{d-2}  e^{\ii (q x \cos \theta_c+q r_\bot \sin\theta_c \cos\phi)} \\
	\nonumber
	&\hspace{5cm}\times \frac{D B_1(\theta_c)\sin^2(\theta_c)}{2q^2(2 A_2(\theta_c) B_1(\theta_c) \nu q^2 + F^{\prime\prime}(\theta_c)(\theta-\theta_c)^2)} \\
	\nonumber
	&\rightarrow\frac{\Gamma\Big(\frac{d}{2}\Big)S_d}{\pi\Gamma\Big(\frac{d-2}{2}\Big)(2\pi)^d}  \int_0^\infty \dd q\, q^{d-1} \int_0^\pi \dd \phi \sin(\phi)^{d-3}  \frac{ D B_1(\theta_c) \sin(\theta_c)^{d}}{F^{\prime\prime}(\theta_c)mq^3}  e^{\ii (q x \cos \theta_c+q r_\bot \sin\theta_c \cos\phi)} \int_{-\infty}^{\infty} \dd  y \frac{1}{y^2+1} \ , \\
	&=\frac{\Gamma\Big(\frac{d}{2}\Big)S_d}{\Gamma\Big(\frac{d-2}{2}\Big)(2\pi)^d}  \int_0^\infty \dd q\, q^{d-1} \int_0^\pi \dd \phi \sin(\phi)^{d-3}  \frac{ D B_1(\theta_c) \sin(\theta_c)^{d}}{F^{\prime\prime}(\theta_c)mq^3}  e^{\ii (q x \cos \theta_c+q r_\bot \sin\theta_c \cos\phi)}\ , 
\end{align}
which becomes exact in the limit of $q\rightarrow 0$, i.e., in the hydrodynamic limit.

The scaling for $C_{LL}$ is obtained completely analogously, such that we obtain,
\begin{equation}
	\chi_L = \chi_\rho = \frac{3-d}{2} \sep z = 4  \sep \zeta = 1 \ .
\end{equation}
This linear-level scaling behaviour is clearly different from that of the Lifshitz point discussed above and, as far as we are aware different from the scaling behaviour of any other known critical point at the linear level, thus highly suggestive that the true scaling behavior, upon incorporating the relevant nonlinear terms into the analysis, will be different as well.

\subsubsection{Critical point $C_{4a}$}

Now we turn to the case $\theta_c=0$. Here the scaling behaviour changes, due to powers of $\sin\theta$ appearing in the integrands of the correlation functions, due to the Jacobian of the angular integration, as well as the density correlation function integrand being proportional to $q_\bot^2$,
\begin{equation}
	C_{\rho\rho} = \frac{\Gamma\Big(\frac{d}{2}\Big)S_d}{\pi\Gamma\Big(\frac{d-2}{2}\Big)(2\pi)^d}  \int_0^\infty q^{d-1}\dd q \int_0^\pi \sin(\phi)^{d-3} \dd \phi\int_0^\pi \dd \theta \sin(\theta)^{d-2}  e^{\ii \bq\cdot\br} \frac{D B_1(\theta)\sin^2(\theta)}{2q^2(A_2(\theta)B_1(\theta)^2+B_2(\theta)^2-2A_2(\theta_c)\nu q^2)}\ .
\end{equation}
The Jacobian factor and the additional factor $\sin^2(\theta)$, regularize the angular integral sufficiently for $d>1$, such that the limit $q\rightarrow 0$ can be taken immediately, by setting $\nu=0$. The integral is convergent if $d>1$, such that the angular integral is a regular function of $q_x x$ and $\bq_\bot\cdot r_\bot$, therefore not changing the scaling behaviour of the correlation function compared to what is expexted in the ordered phase,
\begin{equation}
	\chi_\rho = \frac{2-d}{2}  \sep z = 4  \sep \zeta = 1 \ ,
\end{equation}
besides of course the dynamic exponent $z$.
For $C_{LL}$ we can come to a similar conclusion, except that  the convergence of the angular integral depends on which of the eigenvalues becomes critical. If it is the eigenvalue that can be critical independently from $C_3$, then the numerator also goes to zero as $\theta^2$ and the situation is the same as for the $C_{\rho\rho}$ correlation function. However, if it is the eigenvalue that is equal to $\mu_x$ for $\theta=0$, then to ensure convergence of the integral, the dimension must be larger, i.e., $d>3$. For the case $d < 3$, to extract the scaling dimension we can do the same reparametrization as for $C_{4b}$, i.e.,
\begin{align}
	\nonumber
	C_{LL} &= \frac{\Gamma\Big(\frac{d}{2}\Big)S_d}{\pi\Gamma\Big(\frac{d-2}{2}\Big)(2\pi)^d}  \int_0^\infty q^{d-1}\dd q \int_0^\pi \sin(\phi)^{d-3} \dd \phi\int_0^\pi \dd \theta \sin(\theta)^{d-2}  e^{\ii \bq\cdot\br} \frac{D\Big[\big(\xi^2\cos^2(\theta)+A_2(\theta)\big)B_1(\theta)-2\xi\cos(\theta)B_2(\theta)\Big]}{2q^2(A_2(\theta)\big(B_1(\theta)+\nu q^2\big)^2-B_2(\theta)^2)} \\
	\nonumber
	&= \frac{\Gamma\Big(\frac{d}{2}\Big)S_d}{\pi\Gamma\Big(\frac{d-2}{2}\Big)(2\pi)^d}  \int_0^\infty q^{d-1}\dd q \int_0^\pi \sin(\phi)^{d-3} \dd \phi \frac{(mq)^{d-3}}{2F^{\prime\prime}(0)q^2} e^{\ii q x} D\Big[\big(\xi^2+A_2(0)\big)B_1(0)-2\xi B_2(0)\Big]\\
	&\hspace{5cm}\times\int_{-\infty}^\infty \dd y y^{d-2} \frac{e^{\ii q^2 r_\bot m y \cos\phi} }{y^2 +1} \  \\
	&= \frac{\Gamma\Big(\frac{d}{2}\Big)S_d}{\pi\Gamma\Big(\frac{d-2}{2}\Big)(2\pi)^d}  \int_0^\infty q^{d-1}\dd q \int_0^\pi \sin(\phi)^{d-3} \dd \phi \frac{(mq)^{d-3}}{2F^{\prime\prime}(0)q^2} e^{\ii q x}  D\Big[\big(\xi^2+A_2(0)\big)B_1(0)-2\xi B_2(0)\Big]  \\
	&\hspace{5cm}\times\frac{\Gamma\Big(\frac{3-d}{2}\Big)\Gamma\Big(\frac{1+d}{2}\Big)}{d-1} e^{ - q^2 |m  r_\bot \cos\phi|  } \ ,
\end{align}
from which we can read off that,
\begin{equation}
	\label{eq:crit_2_mf_LL}
	\chi_L = \left\{ \begin{array}{ccc}
		\frac{2-d}{2} & \text{if} & d > 3 \\
		\frac{5-2d}{4} & \text{if} & d < 3 
	\end{array} \right.
	\sep z= 4  \sep \zeta = \left\{ \begin{array}{ccc}
		1 & \text{if} & d > 3 {\rm\ and\ } \mu_x =0 {\rm\ or\ } d > 1 {\rm\ and \ } \mu_x >0 \\
		2 & \text{if} & 3 > d > 1  {\rm\ and\  } \mu_x =0
	\end{array} \right. \ .
\end{equation}
Again, this critical behaviour is novel already at the linear level, as far as we are aware. For $d=3$, we expect logarithmic corrections.

\bibliography{references}